Theoretical study on the core-excited states of the allyl using CVS-icMRCISD method

Qi Song,[1] Junfeng Wu,[1] Wenli Zou,[1] Yibo Lei,[2] and Bingbing Suo[1]

1. Institute of Modern Physics, Northwest University, Xi'an, Shaanxi 710069, China
2. College of Chemistry and Materials Science, Northwest University, Xi'an, Shaanxi 710069, People's Republic of China

**Abstract**

The allyl radical ($C_3H_5$) is a well-characterized hydrocarbon radical, renowned for its pivotal role as an intermediate species in high-energy environments. Its core excited states can elucidate intricate details pertaining to its electronic and structural properties. The core excited states of allyl were studied experimentally using X-ray absorption spectroscopy (XAS), and the primary characteristic peaks were assigned using the MCSCF approach, but not entirely. In this work, the recently developed CVS-icMRCISD scheme was used to simulate the excitation and ionization processes of C's *K*-shell electrons within allyl radicals, cations, and anions, respectively. Our results indicate that the XAS spectrum obtained not merely captured the distinctive peaks associated with allyl radicals, but also encompassed the characteristic peaks pertaining to allyl cations. Meanwhile, unlike manually adjusting the state weights of different electronic states to align with experimental spectral data, we adopt the CVS-icMRCISD scheme, which uses state averaging and produces unbiased results, making it suitable for studying multiple states simultaneously and easy to converge. More importantly, when accounting for the dynamic electron correlation, our results align seamlessly with the experimental XAS. And this congruence underscores the potential of our CVS-icMRCISD as a robust tool for theoretical investigations pertaining to the excitation of inner shell electrons in small molecules.

1. **Introduction**

The allyl radical (C₃H₅), characterized by its π-bonded arrangement of three carbon atoms, exhibits diverse electronic configurations and participates in numerous chemical reactions [1–3], making it a compelling subject for theoretical study [4–7]. Its open-shell electronic configuration, due to an unpaired electron, endows it with high reactivity and versatility in chemical transformations, playing a pivotal role in radical chemistry and organic synthesis [8]. As an exemplary instance of a π-conjugated electron system in its most fundamental form, allyl holds significance in both theoretical understanding and practical applications. Allyl's reactivity facilitates efficient electron transfer, resulting in the formation of allyl cations or anions, which are crucial in various synthetic processes, including cycloaddition reactions for the synthesis of carbocyclic rings and C-C bond formation reactions [9]. Additionally, allyl radicals are important intermediates during combustion processes, particularly in hydrocarbon-rich flames, where they contribute to the synthesis of polycyclic aromatic hydrocarbons (PAHs) and soot [10]. In environments like interstellar space, theoretical investigations into the core-excited states of allyl are essential due to the prevalence of hydrocarbon radicals subjected to high-energy radiation.

To understand the allyl photochemistry involved in enriched ices exposed to X-ray radiation at the molecular level, M. Alagia *et al.* utilized high-resolution synchrotron radiation to investigate the X-ray absorption spectrum (XAS) of allyl radicals within a supersonic helium seed beam at elevated temperatures [11]. They designated four distinctive characteristic peaks pertaining to the allyl radicals, namely bands A to D, using state-average multi-configuration self-consistent field (SA-MCSCF) level *ab initio* calculations. Nevertheless, there exist certain aspects deserving further deliberation, particularly the reassessment of the shoulder peaks and split peak, which is the purpose of this work.

In this work, the core-excited states of allyl radical, cation and anion have been comprehensively simulated using our latest developed CVS-icMRCISD algorithm, which implemented the core-valence separation (CVS) approximation to the multi-reference configuration interaction method with single and double excitations (MRCISD or MRCI for short).[12] The MRCI method, a variational approach, has long been recognized as a benchmark tool for achieving highly precise electronic structure calculations of atoms and molecules. The principal advantage of MRCI is its ability to yield accurate and reliable descriptions of states exhibiting multi-configurational characteristics when an appropriately large basis set and a reasonable reference space are employed [12]. In the MRCI based on graphical unitary group approach (GUGA), the CVS approximation can be achieved through the flexible truncation of the Distinct Row Tableau (DRT). GUGA not only provides a compact way to record configuration state functions (CSFs) of CI calculations within a DRT, but also facilitates the computation of coupling coefficients through a straightforward method.[13–17]

The key point of the CVS approximation is to separate the core excitation configurations from the lower energy valence excitation configurations. Since core orbitals are strongly spatially localized around the corresponding atoms, the energy and spatial localization differences between core and valence orbitals are so significant that the couplings between core- and valence-excited states are very small and can be ignored [18]. According to the GUGA and CVS approximation, the valence-excited configurations represented by Gelfand states with "step 3" applied to all target core orbitals in the extended active space should be eliminated. The implementation of CVS involves the omission of certain nodes and arcs within the sub-DRTs of the extended active space.[15]

The rigorous benchmark calculations for the excited state of K-shell electrons in the first row of elements within small molecules have demonstrated that the CVS-icMRCISD /aug-cc-pVTZ is capable of reproducing core-excited energies that are consistent with the experiments. Specifically, the calculated vertical excitation energy of the 1s electron for the first-row elements exhibits a deviation center of merely −0.09 eV, accompanied by a standard deviation of 0.14 eV [12]. Based on these results, we are confident that CVS-icMRCISD can offers a superior explanation for the core electron excitation spectra of allyl radicals.

2. **Methods and Computational Details**

As mentioned earlier, large basis sets are imperative for investigating the excited states of core electrons within molecular systems. All CVS-icMRCISD calculations were carried out for a range of basis sets, including aug-cc-pVDZ and aug-cc-pVTZ. The effect of relativistic effects was evaluated using the relativistic aug-cc-pVDZ-DK and aug-cc-pVTZ-DK basis sets, incorporating the X2C Hamiltonian to account for scalar relativistic corrections.[19–21] Nevertheless, the core–valence correlation described by these basis sets was found to have a negligible effect on the simulated core-excited spectra. The ground-state equilibrium geometries of the allyl systems (radical, cation and anion) have undergone rigorous optimization within the $C_{2v}$ point group at MCSCF level level of theory with the aug-cc-pVTZ basis set, as this combination was shown to provide accurate structures. The optimized, experimental results[22] and theoretical values from previous works[23] are listed in table 1. All calculations were carried out with the BDF program package[24]. In particular, both the CASSCF and Xi'an-CI modules within BDF were used to perform the core-excited multi-state calculations.

Given the inherent open-shell nature of the allyl radical, its spin multiplicity is 2. However, upon the occurrence of electron donation or acceptance, leading to the formation of the allyl cation or anion, respectively, they exhibit a closed-shell characteristic, resulting in a spin multiplicity of 1. The open-shell ground state electronic configuration and all the orbitals selected in the CVS-MCSCF and CVS-icMRCISD calculation of the allyl radicals can be

described based on the Hartree-Fock molecular orbitals (MO) in $C_{2v}$ point group:

$(1a_1)^2(1b_2)^2(2a_1)^2\text{-}(3a_1)^2(2b_2)^2(4a_1)^2(5a_1)^2(3b_2)^2(4b_2)^2(6a_1)^2(1b_1)^2(1a_2)^1(2b_1)^0$.

or in $C_s$ point group as:

$(1a')^2(2a')^2(3a')^2\text{-}(4a')^2(5a')^2(6a')^2(7a')^2(8a')^2(9a')^2(10a')^2(1a'')^2(2a'')^1(3a'')^0$.

While the ground-state electronic configurations and selected orbital of the allyl cation and anion in the $C_s$ point group as are

$(1a')^2(2a')^2(3a')^2\text{-}(4a')^2(5a')^2(6a')^2(7a')^2(8a')^2(9a')^2(10a')^2(1a'')^2(2a'')^0(3a'')^0$

and

$(1a')^2(2a')^2(3a')^2\text{-}(4a')^2(5a')^2(6a')^2(7a')^2(8a')^2(9a')^2(10a')^2(11a')^0(12a')^0(13a')^0(1a'')^2(2a'')^2(3a'')^0(4a'')^0$, respectively. A total of 13 orbitals were selected for the allyl radical and cation, while 17 orbitals for the anion.

Both the core and frontier orbitals of the allyl radical are depicted in Figure 1. The first three orbitals, 1a', 2a' and 3a' in $C_s$ symmetry, represent the target core orbitals. Specifically, 1a' corresponds to the 1s atomic orbital (AO) localized on the central carbon ($C_C$) atom, whereas 2a' and 3a' are orbitals distributed across the two terminal carbon ($C_T$) atoms. These latter two orbitals arise from the combination of the antisymmetric and symmetric components of the 1s AO of the $C_T$ atom, and delocalize over two $C_T$ atoms. As a crucial step in the subsequent calculation process, it is necessary to localize this pair of orbitals within the A' irreducible representation under $C_s$ symmetry.

The singly occupied molecular orbital (SOMO) of the allyl radical, along with the lowest unoccupied molecular orbital (LUMO) of the allyl cation, are in the A'' irreducible representation of $C_s$ symmetry, while the LUMO of the allyl anion is in the A' irreducible representation. Therefore, a uniform active space was designated for both allyl radical and cation, whereas a distinct and suitable active space was chosen for the anion. The detailed CVS-icMRCISD calculation procedures for the core-excited states of the allyl systems, including the allyl cation, radical and anion, are outlined as follows:

i. Perform a normal MCSCF calculation to optimize the molecular orbitals for the ground state in the $C_s$ point group symmetry.
ii. Followed by a normal MRCI calculations to obtain the energy eigenvalue and corresponding wave-function of the ground state. For allyl radical and cation, a total of 13 orbitals were selected, of which 10 orbitals in the A' irreducible representation were set to be double occupied, and 3 orbitals in the A" irreducible representation were assigned to the active space. For the anion, a total of 17 orbitals were selected, with 10 orbitals in the A' irreducible representation set to be doubly occupied, and the remaining orbitals assigned to the active space.

iii. Read in the molecular orbitals optimized by MCSCF and localize the three core orbitals with the Pipek-Mezey localization approach[25]. After the localization, the 1a' orbital remains localized on the $C_C$ atom, while the 2a' and 3a' orbitals localize on separate $C_T$ atoms, respectively.

iv. Rotate the optimized target core orbitals and appended into the active space, perform a CVS-MCSCF calculation for core-excited states, and generate CVS reference configurations. In this calculation, 9 orbitals in the A' irreducible representation were set to be double occupied.

v. Invoke the single- and double-excitation operators on the CVS-reference configurations and eliminate redundant and all valence excitation configurations.

vi. Perform a CVS-icMRCISD calculation for the core-excited states with the variational approach to obtain the eigenvalues and corresponding eigenfunctions of these states. State-averaged mode with equal weights for each state was employed.

vii. Calculated the transition dipole moment between the ground state and the core-excited state CI wavefunction to determine transition oscillator strengths and relative transition intensities. Degeneracy of the transitions from the two terminal carbon atoms should be taken into account.

3. **Results and Discussion**

**3.1. Structures of Allyl cation, radical and anions**

The Hartree-Fock core orbitals (1a', 2a', 3a') of the allyl systems and the corresponding orbital energies are illustrated in Figure 1. Three obvious features can be identified at a glance. Firstly, the core orbitals 2a' and 3a' in each of allyl systems are nearly degenerate. The core orbitals 2a' and 3a' (localized on the two $C_T$ atoms, respectively) are referred to as the $C_T$ 1s, while the core orbital 1a' (localized on the $C_C$ atom) is referred to as $C_C$ 1s. Secondly, the electronic screening effect results in a consistent increase in the energy of three core orbitals, which is positively correlated with the number of electrons present in the allyl systems [26]. Specifically, the $C_C$ 1s orbital energy of the allyl cation, radical, and anion are measured to be -312.28, -305.83, and -300.22 eV, respectively. Similarly, the $C_T$ 1s orbital energy of the allyl cation, radical, and anion are observed to be -314.24, -305.67, and -298.62 eV, respectively. Thirdly, the energy of $C_T$ 1s orbitals exhibits a steeper increase compared to that of $C_C$ 1s orbitals as the number of electrons in the allyl system increases. In allyl cation, the $C_T$ 1s orbital exhibits the lowest energy level, approximately 1 eV below the energy of the $C_C$ 1s orbitals. Nevertheless, a reversal in the energetic hierarchy of these core orbitals is evident in allyl radical and anion, where the $C_T$ 1s orbital possesses an energy that is 1.5 eV higher than that of the $C_C$ 1s orbitals.

To elucidate the underlying reasons, we employed a range of electronic population analysis methods, including the Mulliken,[27–29] Hirshfeld,[30] and ADCH methods,[31] all seamlessly integrated into the Multiwfn software package[32], to examine the population distribution. The results are presented in table 2. A thorough comparison of the electron population data in table 2 reveals that the increased charge predominantly resides on the two terminal carbon atoms ($C_T$). This observation is consistent with the spatial distribution of the SOMO characteristic of allyl radicals. The shielding effect of the added electrons on the electrons outside the $C_T$ nucleus is more pronounced than that on the $C_C$ nucleus, leading to a significant increase in the energy of the electron orbitals outside the $C_T$ nucleus as the number of electrons rises. Furthermore, table 1 shows that the $C_T$-$C_C$-$C_T$ angles for allyl cations, radicals, and anions are 118.6°, 124.7°, and 135.0°, respectively. The trend indicates that as the number of electrons increases, the $C_T$-$C_C$-$C_T$ angle substantially enlarges. This is due to the fact that the additional electrons are mainly located on the two terminal carbon atoms, and the repulsion between the $C_T$s increases the $C_T$-$C_C$-$C_T$ angle.

**3.2 Vertical ionizations and excitations of *K*-shell electron**

Theoretical computation of core electron vertical ionization potentials (CEVIPs) is essential for interpreting experimental data obtained from X-ray photoelectron spectroscopy. This computation provides valuable insights into the nature of chemical interactions and bond formation within gaseous environments. The calculation of CEVIPs involves assessing the energy difference between the ground state of the neutral molecule and its corresponding core-ionized cation. Table 3 presents the CEVIPs and the lowest lying singly excited energies of the carbon 1s orbital electron for the allyl system, calculated using localized orbitals with the CVS-icMRCISD method combined with the aug-cc-pVTZ basis set. From table 3, there is a clear trend of decreasing CEVIP with the increase in charge on the allyl system. This trend can be observed in the ionization energies of the $C_C$ and $C_T$ 1s orbital electron. For allyl cation, the ionization energies are 297.28 eV and 298.08 eV, respectively. These values decrease to 289.79 eV and 290.26 eV for allyl radical. Finally, for allyl anion, the ionization energies further decrease to 283.83 eV and 280.77 eV, respectively. This consistent reduction in CEVIP reflects the shielding effect of electron charge on the ionization potential of the system's inner-shell electrons. The additional charge reduces the effective nuclear charge experienced by the electrons, lowering the energy required to ionize an electron from the core orbitals. Due to the fact that the additional electrons are mainly concentrated in $C_T$, they have a stronger shielding effect on the core electrons of $C_T$, resulting in a more significant decrease in CEVIP of $C_T$ compared to $C_C$.

### 3.3 Excitation energies and transition moments

The calculated *K*-shell electron excitation energies of carbon within allyl radicals and cations, along with their respective relative transition intensities, have been determined using the CVS-MRCI method in conjunction with the aug-cc-pVTZ basis set. The results are detailed in table 4, and a visual representation of the data is provided in figure 2. Due to the lack of comparative significance in the relative transition intensities of the core electronic excited states of carbon in allyl radicals and cations, the transition intensities associated with band A and band α are chosen as references, setting them to 1, respectively. Electronic transitions with transition oscillator strengths $f_k$ less than 0.01 were ignored.

Our findings confirm the previous conclusion that the distinguishable bands A and C in the experimentally obtained XAS of the allyl radical correspond to electronic transitions from the 1s orbital of $C_T$ to the frontier orbitals (SOMO and LUMO). Meanwhile, bands B and D correspond to electronic transitions from the 1s orbital of $C_C$ to the LUMO, with differences in configuration weights.

However, our labeling of the non-characteristic main peaks in figure 2, specifically the α band at 282.52 eV and the β band at 286.92 eV, differs from previous conclusions. The α and β bands are assigned to the excitation of 1s orbital electrons of carbon in allyl cations. Specifically, the α band corresponds to the electronic transition of allyl cations from $C_T$ to LUMO, configuration weight of about 78%, with a calculated excitation energy of 282.64 eV. The calculated excitation energy for the β band is 286.88 eV. Configuration analysis shows that the β band exhibits significant multi-configuration effects, with corresponding electronic transitions from $C_T$ to LUMO+1 (configuration weight of about 41%) and a double excitation from $C_T$ and HOMO to LUMO (configuration weight of about 40%). As shown in figure 3, the calculated results align completely with the experimental XAS, indicating that the excitation of C 1s orbital electrons in the allyl cation should also be considered when interpreting the experimental XAS spectrum of allyl groups.

Previous studies have held the opinion that band A exhibits a complex vibrational structure, and its electronic transitions are closely related to nuclear dissociation dynamics. The geometric structure of the core excited state pertaining to band A was optimized using TDDFT. The results showed that the $CH_2$ group in allyl twisted about 30° with respect to the molecular plane and the lengths of the two C-H bonds in this group significantly increased from 1.07 Å to 1.13 Å.[11] To verify this conclusion, we used the dihedral angle of $C_T$-$C_C$-$C_T$-H in allyl radicals as the independent variable, adjusting the range from -90° to 90° to scan the potential energy curves of these core electron excited states. In these calculations, the point group symmetry of the allyl radical is reduced to $C_1$, as the rotation of the $CH_2$ fragment reduces the symmetry of the allyl radical structure. Three energy eigenvalues were calculated in the state-average mode, but only the first two were recorded and illustrated in figure 3. From figure 3, it is clear that the excited states of $C_T$ 1s electron still tend to maintain a planar structure. Additionally, it is important to note that the potential barrier for $CH_2$ rotation is relatively low, about 1.0 eV. If the rotation is thermally activated, $CH_2$ can undergo rotation. This conclusion is drawn from the analysis of the constrained potential energy surface results, as our current computational code does not support structural optimization and frequency analysis of core excited states. Therefore, further experiments are needed to eliminate accidental factors and provide additional evidence in the future.

To validate the results of CVS-icMRCISD for the study of core excited states in allyl systems and to assess the efficacy of various schemes in addressing core excited states, we also employed additional methods including CVS-icMRCSID+Q (with Davidson correction), the multi-state *N*-electron valence state second-order perturbation theory with core-valence separation approximation (CVS-MS-NEVPT2) and the static-dynamic-static multi-state multi-reference second-order perturbation theory with core valence separation approximation and Davidson correction (CVS-SDSPT2+Q3) to study the excited states of the C 1s orbital electrons in allyl systems. The results are presented in Table 5. To uphold consistency and ensure unbiased comparisons, all methods adopted identical active spaces and the same basis set (aug-cc-pVTZ).

The deviation between theoretical and experimental values is indicated in parentheses. Table 5 clearly shows that the CVS-icMRCISD calculated results align closely with the experimental values, with an average absolute deviation of only 0.24 eV. After considering the Davidson correction, the theoretical values are slightly lower than the experimental ones, with an average absolute deviation of 0.43 eV. Additionally, the Davidson correction reduces the overall excitation energy by approximately -0.61 eV. The CVS-MCSCF method, lacking dynamic electron correlation consideration, exhibits a high average absolute deviation of 2.65 eV from the experimental results, nearly five times greater than that of CVS-icMRCISD. To align the theoretical spectrum with the experimental one, energy shift factors or artificial adjustment of state weights would be required, which is imprecise and impractical. Perturbation methods such as CVS-NEVPT2 and CVS-SDSPT2+Q3, which partially account for dynamic electron correlation, show average absolute deviations of 0.71 and 0.58 eV, respectively, between the calculated and experimental values. However, due to the instability inherent in perturbation theory, the average absolute deviations for bands B and D are notably high, at 1.47 and 1.59 eV, respectively.

The impact of various basis sets on the core excited states calculations have also been evaluated. Here, we used CVS-icMRCISD, incorporating the aug-CC-pVDZ, aug-CC-pVTZ, aug-CC-pVDZ-DK, and aug-CC-pVTZ-DK basis sets to meticulously calculate and compare the carbon 1s orbital electron excited states of the allyl radical. Notably, the utilization of the latter two basis sets, coupled with the X2C Hamiltonian, enabled us to effectively account for the influence of scalar relativistic effects on the core electron excited states. The results were depicted in Figure 4. The results confirm that our developed CVS-icMRCISD method can obtain accurate results directly, without any energy shift, when using a large basis sets such as aug-cc-pVTZ. When utilizing a smaller basis set like aug-cc-pVDZ, the calculated characteristic peak shapes and the relative positions between these characteristic peaks align well with experimental values. However, there is a minor discrepancy in the exact positions of the characteristic peaks. This can be rectified by introducing an appropriate energy shift to better match the experimental spectra. We also employed scalar relativistic basis sets, such as aug-cc-pVDZ-DK and aug-cc-pVTZ-DK to calculate these core electron excited states. The findings revealed that, taking into account relativistic effects, the average core orbital energy corrections is -0.1 eV for the carbon 1s orbital, which is consistent with our earlier findings[12].

4. **Conclusions**

In this work, we utilized our recently developed CVS-icMRCISD scheme to simulate the excitation and ionization processes of carbon's *K*-shell electrons within allyl radicals, cations, and anions. The results indicate that as we progress from allyl cations to allyl radicals and then to allyl anions, the increasing number of electrons in the system leads to enhanced electron shielding effects. Consequently, there is a continuous increase in the energy of the 1s orbitals of the three carbon atoms in the allyl compounds. Population analysis reveals that the obtained charges are predominantly distributed on the terminal carbon ($C_T$), resulting in a reversal of the 1s orbital order of the central carbon ($C_c$) and $C_T$. This reversal ultimately affects the excitation energy and vertical ionization potential of the 1s orbital electrons in the allyl systems.

We calculated the 1s electron excitation energy of carbon in the allyl radical using the CVS-icMRCISD method combined with the aug-cc-pVTZ basis set and annotated four spectral features on the experimental X-ray absorption spectrum of the allyl radical. Our attribution of the electronic states corresponding to the four characteristic peaks aligns with previous conclusions; however, our results indicate that companion peaks α and β should be attributed to the excited states of the 1s orbital electrons of carbon which differs from the assignment of the earlier study.

References:

(1) Gordon, A. S.; Smith, S. R.; McNesby, J. R. Some Reactions of the Allyl Radical. *J. Am. Chem. Soc.* 1959, *81* (19), 5059–5061. https://doi.org/10.1021/ja01528a012.

(2) Lee, J.; Bozzelli, J. W. Thermochemical and Kinetic Analysis of the Allyl Radical with $O_2$ Reaction System. *Proceedings of the Combustion Institute* 2005, *30* (1), 1015–1022. https://doi.org/10.1016/j.proci.2004.08.092.

(3) Miller, J. A.; Klippenstein, S. J.; Georgievskii, Y.; Harding, L. B.; Allen, W. D.; Simmonett, A. C. Reactions between Resonance-Stabilized Radicals: Propargyl + Allyl. *J. Phys. Chem. A* 2010, *114* (14), 4881–4890. https://doi.org/10.1021/jp910604b.

(4) Korth, H.-G.; Trill, H.; Sustmann, R. [1-2H]-Allyl Radical: Barrier to Rotation and Allyl Delocalization Energy. *J. Am. Chem. Soc.* 1981, *103* (15), 4483–4489. https://doi.org/10.1021/ja00405a032.

(5) Fischer, I.; Chen, P. Allyl-A Model System for the Chemical Dynamics of Radicals. *J. Phys. Chem. A* 2002, *106* (17), 4291–4300. https://doi.org/10.1021/jp013708o.

(6) Shahu, M. CASSCF Study into the Mechanism for Predissociation of the Allyl Radical. *Int. J. Quantum Chem.* 2006, *106* (2), 501–506. https://doi.org/10.1002/qua.20757.

(7) Oliva, J. M.; Gerratt, J.; Cooper, D. L.; Karadakov, P. B.; Raimondi, M. Study of the Electronic States of the Allyl Radical Using Spin-Coupled Valence Bond Theory. *The Journal of Chemical Physics* 1997, *106* (9), 3663–3672. https://doi.org/10.1063/1.473460.

(8) Wang, L.; Wang, L.; Li, M.; Chong, Q.; Meng, F. Cobalt-Catalyzed Diastereo- and Enantioselective Reductive Allyl Additions to Aldehydes with Allylic Alcohol Derivatives via Allyl Radical Intermediates. *J. Am. Chem. Soc.* 2021, *143* (32), 12755–12765. https://doi.org/10.1021/jacs.1c05690.

(9) Hoffmann, H. M. R. Syntheses of Seven- and Five-Membered Rings from Allyl Cations. *Angew. Chem. Int. Ed. Engl.* 1973, *12* (10), 819–835. https://doi.org/10.1002/anie.197308191.

(10) Marinov, N. M.; Pitz, W. J.; Westbrook, C. K.; Vincitore, A. M.; Castaldi, M. J.; Senkan, S. M.; Melius, C. F. Aromatic and Polycyclic Aromatic Hydrocarbon Formation in a Laminar Premixed N-Butane Flame. *Combustion and Flame* 1998, *114* (1–2), 192–213. https://doi.org/10.1016/S0010-2180(97)00275-7.

(11) Alagia, M.; Bodo, E.; Decleva, P.; Falcinelli, S.; Ponzi, A.; Richter, R.; Stranges, S. The Soft X-Ray Absorption Spectrum of the Allyl Free Radical. *Phys. Chem. Chem. Phys.* 2013, *15* (4), 1310–1318. https://doi.org/10.1039/C2CP43466K.

(12) Song, Q.; Liu, B.; Wu, J.; Zou, W.; Wang, Y.; Suo, B.; Lei, Y. GUGA-Based MRCI Approach with Core-Valence Separation Approximation (CVS) for the Calculation of the Core-Excited States of Molecules. *The Journal of Chemical Physics* 2024, *160* (9),

094114. https://doi.org/10.1063/5.0189443.

(13) Shavitt, I. Graph Theoretical Concepts for the Unitary Group Approach to the Many-Electron Correlation Problem. *Int. J. Quantum Chem.* 2009, *12* (S11), 131–148. https://doi.org/10.1002/qua.560120819.

(14) Shavitt, I. Matrix Element Evaluation in the Unitary Group Approach to the Electron Correlation Problem. *Int. J. Quantum Chem.* 2009, *14* (S12), 5–32. https://doi.org/10.1002/qua.560140803.

(15) Shamasundar, K. R.; Knizia, G.; Werner, H.-J. A New Internally Contracted Multi-Reference Configuration Interaction Method. *The Journal of Chemical Physics* 2011, *135* (5), 054101. https://doi.org/10.1063/1.3609809.

(16) Wang, Y.; Han, H.; Lei, Y.; Suo, B.; Zhu, H.; Song, Q.; Wen, Z. New Schemes for Internally Contracted Multi-Reference Configuration Interaction. *The Journal of Chemical Physics* 2014, *141* (16), 164114. https://doi.org/10.1063/1.4898156.

(17) Werner, H.-J.; Knowles, P. J. An Efficient Internally Contracted Multiconfiguration–Reference Configuration Interaction Method. *The Journal of Chemical Physics* 1988, *89* (9), 5803–5814. https://doi.org/10.1063/1.455556.

(18) Norman, P.; Dreuw, A. Simulating X-Ray Spectroscopies and Calculating Core-Excited States of Molecules. *Chem. Rev.* 2018, *118* (15), 7208–7248. https://doi.org/10.1021/acs.chemrev.8b00156.

(19) Cao, Z.; Li, Z.; Wang, F.; Liu, W. Combining the Spin-Separated Exact Two-Component Relativistic Hamiltonian with the Equation-of-Motion Coupled-Cluster Method for the Treatment of Spin–Orbit Splittings of Light and Heavy Elements. *Phys. Chem. Chem. Phys.* 2017, *19* (5), 3713–3721. https://doi.org/10.1039/C6CP07588F.

(20) Reiher, M. Douglas–Kroll–Hess Theory: A Relativistic Electrons-Only Theory for Chemistry. *Theor Chem Acc* 2006, *116* (1–3), 241–252. https://doi.org/10.1007/s00214-005-0003-2.

(21) Jansen, G.; Hess, B. A. Revision of the Douglas-Kroll Transformation. *Phys. Rev. A* 1989, *39* (11), 6016–6017. https://doi.org/10.1103/PhysRevA.39.6016.

(22) Vajda, E.; Tremmel, J.; Rozsondai, B.; Hargittai, I.; Maltsev, A. K.; Kagramanov, N. D.; Nefedov, O. M. Molecular Structure of Allyl Radical from Electron Diffraction. *J. Am. Chem. Soc.* 1986, *108* (15), 4352–4353. https://doi.org/10.1021/ja00275a020.

(23) Linares, M.; Humbel, S.; Braïda, B. The Nature of Resonance in Allyl Ions and Radical. *J. Phys. Chem. A* 2008, *112* (50), 13249–13255. https://doi.org/10.1021/jp8038169.

(24) Zhang, Y.; Suo, B.; Wang, Z.; Zhang, N.; Li, Z.; Lei, Y.; Zou, W.; Gao, J.; Peng, D.; Pu, Z.; Xiao, Y.; Sun, Q.; Wang, F.; Ma, Y.; Wang, X.; Guo, Y.; Liu, W. BDF: A Relativistic Electronic Structure Program Package. *The Journal of Chemical Physics* 2020, *152* (6), 064113. https://doi.org/10.1063/1.5143173.

(25) Pipek, J.; Mezey, P. G. A Fast Intrinsic Localization Procedure Applicable for *a b i n i t i o* and Semiempirical Linear Combination of Atomic Orbital Wave Functions. *The Journal of Chemical Physics* 1989, *90* (9), 4916–4926. https://doi.org/10.1063/1.456588.

(26) Linares, M.; Stafström, S.; Rinkevicius, Z.; Ågren, H.; Norman, P. Complex Polarization Propagator Approach in the Restricted Open-Shell, Self-Consistent Field Approximation: The Near *K*-Edge X-Ray Absorption Fine Structure Spectra of Allyl and Copper Phthalocyanine. *J. Phys. Chem. B* 2011, *115* (18), 5096–5102. https://doi.org/10.1021/jp103506g.

(27) Mulliken, R. S. Electronic Population Analysis on LCAO–MO Molecular Wave Functions. I. *The Journal of Chemical Physics* 1955, *23* (10), 1833–1840. https://doi.org/10.1063/1.1740588.

(28) Mulliken, R. S. Electronic Population Analysis on LCAO–MO Molecular Wave Functions. II. Overlap Populations, Bond Orders, and Covalent Bond Energies. *The Journal of Chemical Physics* 1955, *23* (10), 1841–1846. https://doi.org/10.1063/1.1740589.

(29) Mulliken, R. S. Electronic Population Analysis on LCAO-MO Molecular Wave Functions. III. Effects of Hybridization on Overlap and Gross AO Populations. *The Journal of Chemical Physics* 1955, *23* (12), 2338–2342. https://doi.org/10.1063/1.1741876.

(30) Hirshfeld, F. L. Bonded-Atom Fragments for Describing Molecular Charge Densities. *Theoret. Chim. Acta* 1977, *44* (2), 129–138. https://doi.org/10.1007/BF00549096.

(31) Lu, T.; Chen, F. ATOMIC DIPOLE MOMENT CORRECTED HIRSHFELD POPULATION METHOD. *J. Theor. Comput. Chem.* 2012, *11* (01), 163–183. https://doi.org/10.1142/S0219633612500113.

(32) Lu, T.; Chen, F. Multiwfn: A Multifunctional Wavefunction Analyzer. *J Comput Chem* 2012, *33* (5), 580–592. https://doi.org/10.1002/jcc.22885.